\documentclass[10pt,conference]{IEEEtran}
\usepackage{epsfig,rotating,setspace,latexsym,amsmath,epsf,amssymb,amsfonts,bm,theorem,cite,enumerate,longtable,accents,float,physics}
\usepackage{algorithm,algorithmic,graphicx,epsf,authblk,epstopdf,url,xcolor, soul,multirow,bbm}
\usepackage{mathtools,comment}
\usepackage[center]{qtree}
\usepackage{tree-dvips}
\usepackage[linguistics]{forest }

\newtheorem{theorem}{Theorem}
\newtheorem{corollary}{Corollary}
\newtheorem{definition}{Definition}
\newtheorem{remark}{Remark}
\newtheorem{lemma}{Lemma}

\newenvironment{Proof}[1]{\medskip\par\noindent{\bf Proof:\,}\,#1}{{\mbox{\,$\blacksquare$}\par}}

\IEEEoverridecommandlockouts

\allowdisplaybreaks

\title{Quantum Private Membership Aggregation}
\author{Alptug Aytekin \qquad Mohamed Nomeir \qquad Sennur Ulukus\\
	\normalsize Department of Electrical and Computer Engineering\\
	\normalsize University of Maryland, College Park, MD 20742 \\
	\normalsize \emph{aaytekin@umd.edu} \qquad \emph{mnomeir@umd.edu} \qquad \emph{ulukus@umd.edu}}
 
\begin{document}

\maketitle

\begin{abstract}
We consider the problem of private set membership aggregation of $N$ parties by using an entangled quantum state. In this setting, the $N$ parties, which share an entangled state, aim to \emph{privately} know the number of times each element (message) is repeated among the $N$ parties, with respect to a universal set $\mathcal{K}$. This problem has applications in private comparison, ranking, voting, etc. We propose an encoding algorithm that maps the classical information into distinguishable quantum states, along with a decoding algorithm that exploits the distinguishability of the mapped states. The proposed scheme can also be used to calculate the $N$ party private summation modulo $P$.
\end{abstract}

\section{Introduction}

Seemingly superior computational powers of quantum computers pose a threat to classical security and privacy algorithms as shown in \cite{shor}. One way to combat this prevailing threat is to design quantum algorithms. Ever since the first quantum key distribution protocol of \cite{bb84}, quantum cryptography, a field which uses quantum mechanics to carry out classical cryptographic tasks, has become a very active research area. A few of the many important sub-fields of quantum cryptography are quantum key distribution, quantum secret sharing, and quantum multiparty computation. 

\begin{figure}[t]
    \centering
    \includegraphics[width=0.8\columnwidth]{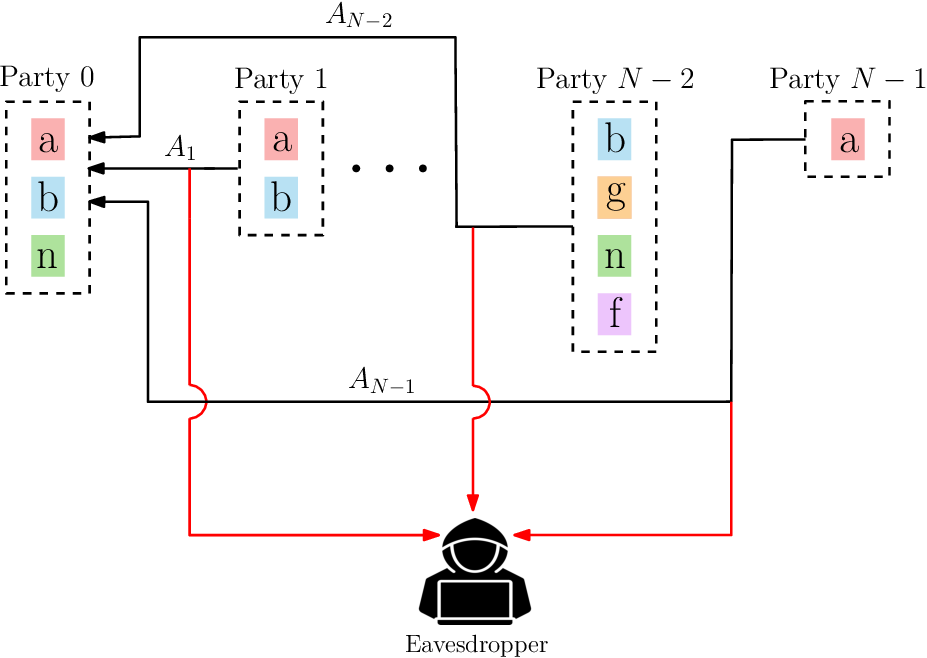}
    \caption{$N$ parties wish to aggregate membership of each element from a universal alphabet in their individual sets privately among themselves, and securely against an external eavesdropper, using an entangled quantum state.}
    \label{fig1}
    \vspace*{-0.4cm}
\end{figure}

As a specific problem in quantum multiparty computation, we consider the problem of quantum private membership aggregation (QPMA). In this problem, there are $N$ parties, each of which having a subset of elements from a universal set. These parties wish to learn the frequency of the occurrence of each element in the universal set while hiding their own subsets from other parties as much as possible, i.e., even after having learned the frequency of all elements in the universal set, any guess about the subset possessed by any party should be equivalent to blind estimation. Further, the parties want to accomplish this in the presence of an eavesdropper which is wiretapping all the communication links. This problem has applications in many settings, such as confidence voting, ranking, and so on, which should ideally be carried out in anonymity for better results. To achieve their goals, parties select one of their own as the leader party which then computes the desired result by using a quantum algorithm. Fig.~\ref{fig1} shows the system model for this problem, with elements from the English alphabet and Party $0$ acting as the leader party.

Classically, various versions of this problem have been studied: \cite{pinkas} introduced the private set intersection (PSI) problem, in which the parties want to learn the elements that exist in all parties, without leaking any further information about their own sets. \cite{zhusheng2021} found the information-theoretic capacity of 2-party PSI; and \cite{zhusheng2022} generalized the scheme to $N$-party PSI. \cite{elkordy} studied a similar problem, in which the parties want to learn whether a certain element occurs at least $K$ times in the parties or not, called the $K$-PSI problem. \cite{mopma} then extended the $K$-PSI problem based on the existing private information retrieval (PIR) schemes \cite{JafarPIR, UlukusPIRLC} to carry out the aggregation privately. We note that a key difference between \cite{mopma} and the current paper is that, in \cite{mopma}, the user wants to learn the frequency of a certain element without leaking the identity of the chosen element, whereas here, the goal is to find the frequency of all elements, that is, there is no privacy requirement on the element whose frequency is being checked. Further, in the current paper, each party possesses only one database as opposed to \cite{mopma} where each party possesses multiple replicated databases.

Even though private quantum summation and aggregation schemes are common in the literature, to our knowledge, ours is the first paper to use an aggregation scheme on set membership function with $N$ parties.\footnote{We remark that even though \cite{novelpsi} does not explicitly state their problem as a PMA problem, it computes the intersection and the union of two sets, which is exactly equivalent to PMA in the two-party case (i.e, $N=2$ case).} Further, it is the first in the way information is encoded and decoded. In \cite{dlevel, bellstates, phaseshifting, singlephotons,unbiased,summul,protocols}, sequential-operating algorithms are used in which party $i$ encodes its information on a set of quantum states, and sends the states to party $i+1$, until a leader party or a trusted third party which has initiated the scheme, compares the initial and the final sets to find the result. Of these, \cite{phaseshifting} is similar to our paper in using Sylvester clock matrix to encode the information. There are parallel-operating algorithms such as \cite{hayashi, aggregating, zhang17, qftsum, efficient, generalizedprot} that make use of quantum states to create a set of shared common randomness that adds up to $0$, which masks their announcement of membership function to the leader party, and then the leader party carries out the summation. What differentiates our scheme from these schemes in the literature is the fact it is a parallel-operating coding scheme that encodes the aggregation results to distinguishable quantum states and decodes thereafter, as opposed to creating ``modulo-zero-sum randomness'' as \cite{hayashi} calls it.

\section{Preliminaries and Notation}
Let $[n]\coloneqq\{0,\ldots,n-1\}$. To indicate the powerset of a set $\mathcal{X}$, $2^\mathcal{X}$ is used. $\mathbf{1}[A]$ is used to denote the indicator function of the event $A$. $e_i^D$ is used to denote the $D$-dimensional vector whose elements are all $0$ except a $1$ at the $i$th index. Whenever the dimensionality of this vector is clear from the context, the superscript is dropped for brevity. $I_k$ denotes the identity matrix of size $k\times k$. $(A)_{ij}$ denotes the $(i,j)$th element of a matrix $A$. All logarithms are taken with respect to base 2.

\begin{definition}[Partial trace operation]
    Let $\{A_i\}_i \in \mathcal{V}_A$ and $\{B_i\}_i \in \mathcal{V}_B$. For the composite matrix $\sum_i A_i \otimes B_i \in \mathcal{V}_A \otimes\mathcal{V}_B$, partial trace operation with respect to a given subsystem is, 
    \begin{align}
        tr_A\left(\sum_i A_i \times B_i\right) &\coloneqq\sum_i tr(A_i) B_i, \\
        tr_B\left(\sum_i A_i \times B_i\right)&\coloneqq\sum_i trA_i tr(B_i).
    \end{align}
\end{definition}

The following definitions follow from \cite{nielsenchuang}.

\begin{definition}[Density matrices]
\label{densitymat}
    Density matrix of the quantum system $A$ which is in the state $\ket{\psi_j}$ with probability $p_j$,
    \begin{align}
        \rho_A \coloneqq \sum_j p_j \ket{\psi_j}\bra{\psi_j},
    \end{align}
    where $p_j \geq 0$, $\sum_j p_j =1$.
\end{definition}

We note here that when $p_j=1$ for some $j$, the density matrix reduces to $\rho=\ket{\psi_j}\bra{\psi_j}$. This is known as a pure state. For brevity, in those cases, instead of $\rho$, $\ket{\psi_j}$ will be used to indicate it.

\begin{definition}[Von Neumann entropy]
    For a density matrix $\rho$ describing the system $A$, Von Neumann entropy is,
    \begin{align}
        S(\rho)=S(A)\coloneqq -tr(\rho \log\rho)=H(\Lambda),
    \end{align}
    where $tr(\cdot)$ is the trace operator, $\Lambda$ is the set of the eigenvalues of $\rho$, and $H(\cdot)$ is the Shannon entropy.
\end{definition}

\begin{definition}[Quantum conditional entropy]
    The conditio-nal entropy of a quantum system $A$ with respect to a quantum system $B$ is,
    \begin{align}
        S(A|B) \coloneqq S(A,B)-S(B).
    \end{align}
    \end{definition}
\begin{definition}[Quantum mutual information]
    The quantum mutual information between two quantum systems $A$, $B$ is,
    \begin{align}
        S(A;B)&\coloneqq S(A)+S(B)-S(A,B)\\
        &=S(A)-S(A|B).
    \end{align}
\end{definition}

\begin{definition}[Quantum operation]
    A quantum operation $\varepsilon$ is a linear, completely-positive map of density matrices.
\end{definition} 

Finally, we use the following simple lemmas that are stated without proofs.

\begin{lemma}
\label{separate}
    Let A and B share a quantum system $\rho$. If $\rho=\rho_A \otimes \rho_B$, where $\rho_i=tr_j(\rho)$, then, $S(A;B)=0$.
\end{lemma}

\begin{lemma}
   Let $\varepsilon(\rho)\coloneqq U\rho U^\dagger$, where $U$ is a unitary matrix. Then, $\varepsilon$ is a linear completely positive trace-preserving, CPTP, map.
\end{lemma}

Let $\mathcal{H}$ be a $d$-dimensional Hilbert space. Let $\{\ket{i}\}_0^{d-1}$ be a basis for it. Let $\omega_d\coloneqq e^{\sqrt{-1}\frac{2\pi}{d}}$. Then, let $\mathsf{Z}_d\coloneqq \sum_{k=0}^{d-1}\omega^k_d\ket{k}\bra{k}$. This is known as the Sylvester clock matrix \cite{sylvester}. Similarly, when the dimensionality is to be understood from the context, the subscript will be dropped.

A projective-valued measurement (PVM) consists of the set $\{P_i\}_i$ such that $P_i^2=P_i$ and $\sum_iP_i=I_d$ where $d$ is the dimension of the underlying Hilbert space $\mathcal{H}$. For a density matrix $\rho$, the result of the PVM $\{P_i\}_i$ is $i$ with probability $tr(P_i\rho)$. For a pure state $\rho=\ket{\xi}\bra{\xi}$, this is simply $\bra{\xi}P_i\ket{\xi}$.
\begin{lemma}
\label{meas}
    An orthonormal basis for the Hilbert space $\mathcal{H}$ constitutes a PVM.
\end{lemma}

Similarly, a partial PVM is a PVM for a subsystem of $\mathcal{H}$, denoted by $\mathcal{H}_a$ with dimension $d_a$. That is, the partial PVM $\{P_{a,i}\}_i$ should satisfy $P_{a,i}^2=P_{a,i}$ and $\sum_iP_{a,i}=I_{d_a}$.

\section{Problem Formulation}\label{problem_formulation}
In this problem, there is a universal set $\mathcal{K}$ with $|\mathcal{K}|=K$. There are $N$ parties, each with a set $\mathcal{N}_i \subseteq \mathcal{K}$ for $0 \leq i \leq N-1$. We note that the probability distribution for realizations of $\mathcal{N}_i$ is not important for the achievable scheme.

Let $\mathbb{F}$ be a field. There exists a bijective map $f$ from $\mathcal{K}$ to $[K]$. Define $g:2^{[K]}\to \mathbb{F}^K:\mathcal{A} \mapsto \sum_{i=0}^{K-1}e_i\mathbf{1}[i \in \mathcal{A}]$. Notice that g is an injection. Thus, the composite map $g\circ f$ is an injection as well so that it is possible to find $\mathcal{N}_i$ from $(g\circ f)(\mathcal{N}_i)$. $(g\circ f)(\mathcal{N}_i)$ is commonly called the incidence vector of the subset $\mathcal{N}_i$ of $\mathcal{K}$. Let $E_{[N]}$ denote the $N$-tuple of incidence vectors of all parties.\footnote{Here, $N$-tuple is used instead of the set to account for the possible repetitions.} The image of $g\circ f$ under a singleton is called the ``set membership output.'' Each party stores its incidence vector $E_i\coloneqq(g\circ f)(\mathcal{N}_i)$ in one server. 

Without loss of generality, any party can be chosen as the leading party. Let the leader be the $L$th party, $0 \leq L \leq N-1$.

In this work, to have privacy/security against an eavesdropper, we allow that party $i$ and the leader party share common randomness $U_i$ prior to the initialization of the scheme. Since the common randomness is shared and the leader party does not previously know the incidence vector $E_i$, we have
\begin{align}
    I(E_i;U_i)=0, \qquad i \in [N].
\end{align}

Before the scheme starts, all parties share a quantum system $\rho_0$ in the composite Hilbert space $\otimes_{k=0}^{N-1}\mathcal{H}_k$. Here, $\mathcal{H}_k$ is the Hilbert space for the quantum system of the $k$th party. When the scheme starts, the $i$th party encodes its information using the mapping $Enc_i$ and then shares its part of the quantum system with the leader party. The quantum system transmitted is called the answer of the $i$th party, denoted with $A_i$.

The answer of the $i$th party depends only on its randomness and its incidence vector, that is,
\begin{align}
    S(A_i|E_i, U_i)=0.
\end{align}

After each party has sent their answers, the leader party will have $\rho_f=(Enc_1 \otimes \ldots \otimes Enc_N)(\rho_0)$. This quantum system will be denoted by $A_{[N]}$.

The leader party should be able to decode the frequency of occurrence of each element in $\mathcal{K}$ by combining the answers,
\begin{align}
    [\mathrm{correctness}]\quad S\left(\sum_{i=1}^{N}E_i \Big| A_{[N]}\right)=0.
\end{align}
This decoding is done by applying a $Dec_L$ map to $\rho_f$.

All parties are considered to be semi-honest (i.e., honest but curious); all parties act according to the described scheme, yet they are curious to learn as much as possible from what they get. Privacy of the parties except the leader party require, 
\begin{align}
 [\mathrm{privacy}] \quad   S(E_{[N]};A_{[N]},U_{[N]})=I&\left(E_{[N]};\sum_{i=1}^{N}E_i\right),
\end{align}
so that the leader party should not be able to learn anything more specific about the incidence vectors than what it could infer after learning $\sum_{i=1}^{N}E_i$.

Moreover, if the quantum channel between party $i$ and the leader party has been wiretapped by a passive eavesdropper, no information about the incidence vectors should leak, i.e., 
\begin{align}
    [\mathrm{security}] \quad S(E_i;A_i)=0,0\leq i \leq N-1, i \neq L.
\end{align}

The download cost of the system is defined with respect to the leader party. From the leader's perspective, it will download $\log\left(\otimes_{k \in [N]\backslash\{L\}}\text{dim}(\mathcal{H}_k)\right)$ qudits.

\section{Main Results}\label{main_results}

\begin{theorem}
\label{main}
    Let the optimal download cost of the quantum private membership aggregation (QPMA) problem be $D^*$. Then,
    \begin{align}
        D^* \leq (N-1)K\log P^*,
    \end{align}
    where $N$ is the number of parties and $P^*$ is the smallest prime number that is larger than $N$.
\end{theorem}

\begin{remark}
    It might seem that the dependence of the download cost in Theorem~\ref{main} on $K$ is bad, but it should be noted that, at the end of the scheme, the frequency of occurrence of all elements in the set $\mathcal{K}$ is found. Thus, on average, per element in the set, the download cost is $(N-1)\log P^*$.
\end{remark}

\section{Proposed Scheme}\label{proposed schemes}
We first give the general scheme in Subsection~\ref{genscheme}, and then present a representative example in Subsection~\ref{repexample}.

\subsection{General Scheme}\label{genscheme}
Let $P$ be the smallest prime number that is at least equal to $N$. Let $\mathbb{F}_P\coloneqq \mathbb{Z}/P\mathbb{Z}=\mathbb{Z}_P$ be the underlying field for the operations on classical information.

Let $\mathcal{H}$ be a $P$-dimensional Hilbert space with the $\{\ket{i}\}_{0}^{P-1}$ basis (a.k.a. computational or $\mathsf{Z}$ basis). Let $\ket{\psi}=\frac{1}{\sqrt{P}}\sum_{k=0}^{P-1}\underbrace{\ket{k\ldots k}}_{\text{$N$ times}}$, and  $\ket{\phi_m}=\frac{1}{\sqrt{P}}\sum_{k=0}^{P-1}\omega_P^{mk}\underbrace{\ket{k\ldots k}}_{\text{$N$ times}}$, where $mk$ is the multiplication in $\mathbb{F}_P$. Also notice that $\ket{\psi}=\ket{\phi_0}$. 

Let $\ket{\psi}^{\otimes K}=\underbrace{\ket{\psi}\otimes\ldots\otimes\ket{\psi}}_{\text{$K$ times}}$. 
The $l$th party has the qudits at $(l+mN)$th locations where $m \in [K]$. That is, to aggregate the occurrence of $ x \in \mathcal{K}$, the parties share an entangled state $\ket{\psi}$ with the $l$th party having the $l$th qudit, for all $x \in \mathcal{K}$.

The $l$th party shares $U_i$ which is uniformly distributed in $\mathbb{F}_P^K$ with the leader. The encoding of the incidence vector to qudits is done by applying the Sylvester clock matrix $\mathsf{Z}_P$, also known as phase operation. The final state after  encoding is,
\begin{align}
    &\ket{\psi_e}=\otimes_{l=0}^{K-1}(\otimes_{k=0}^{N-1}(\mathsf{Z}^{(U_k+E_k)_l}))\ket{\psi}^{\otimes K}\\
    &=(\otimes_{k=0}^{N-1}\mathsf{Z}^{(U_k+E_k)_0})\ket{\psi}\otimes\ldots\otimes(\otimes_{k=0}^{N-1}\mathsf{Z}^{(U_k+E_k)_{K-1}})\ket{\psi}
\end{align}

\begin{remark} \label{leaderno}
    In the encoding stage, since the leader party does not send anything to itself, there is no need to be concerned with the eavesdropper. Thus, the leader party does not apply any shared randomness in its encoding. Moreover, since technically the leader party already knows its set membership results, it does not have to do any encoding at all.
\end{remark}

After this encoding is done, the parties send their quantum systems to the leader party. Then, the leader uses the shared randomness information and does the following,
\begin{align}
\label{decode_step}
        &\ket{\psi_f}=\otimes_{l=0}^{K-1}(\otimes_{k=0}^{N-1}(\mathsf{Z}^{-(U_k)_{l}}))\ket{\psi_e}\\
        \label{sep}&=(\otimes_{k=0}^{N-1}\mathsf{Z}^{(E_k)_0})\ket{\psi}\otimes\ldots\otimes(\otimes_{k=0}^{N-1}\mathsf{Z}^{(E_k)_{K-1}})\ket{\psi}    
\end{align}

Looking at \eqref{sep}, the final state $\ket{\psi_f}$ is effectively separated into $K$ parts each with $N$ qudits. These parts can then be dealt with individually as they form a tensor product state. The motivation of how the individuals parts do not contain any information about each other can be seen from Lemma~\ref{separate}. To decode the information, the leader applies a partial PVM to $N$ qudits to utilize the $K$-separation of $\ket{\psi_f}$ mentioned above. The requirement for the PVM set is that it should include $\ket{\phi_m}\bra{\phi_m}$ for $m \in [N]$. The output of partial PVM at the $l$th location of $K$-separation then gives the membership aggregation result of the element in $\mathcal{K}$ corresponding to $l$. If the measurement gives $\ket{\phi_m}$, then the aggregation result is $m$, whereas if any other output is read, it indicates at least one party is Byzantine. As shown in the scheme, the download cost from each party is $K\log P$. Thus, the leader downloads from $(N-1)$ parties, so that the download cost of the scheme is $(N-1)K\log P$ as stated in Theorem~\ref{main}.

\begin{remark}
    The $\log P$ in the download cost appears since a logarithm with base 2 is being used. The leader party in fact downloads $(N-1)K$ qudits. Thus, one might have actually suggested using $(N-1)K$ as the download cost. However, by using the logarithm with a fixed base, we see that the low download cost will also have to induce a lower field size on $\mathbb{F}_P$, which is a merit since realizing qudits in increasing field sizes can be tricky.
\end{remark}

\subsection{Representative Example}\label{repexample}

Let there be $N=3$ parties, with the universal alphabet $\mathcal{K}=\{a,b,c,d\}$. Let the $0$th, $1$st and $2$nd parties have the sets $\mathcal{N}_0=\{a,c\}$,  $\mathcal{N}_1=\{a,b,c\}$ and $\mathcal{N}_2=\{c\}$, respectively. Then, the incidence vectors are $E_0=(1,0,1,0)$, $E_1=(1,1,1,0)$, and $E_2=(0,0,1,0)$. 

Then, choose $P=3=N$. Thus, a 3-dimensional Hilbert space $\mathcal{H}$ with basis $\{\ket{0},\ket{1},\ket{2}\}$. Thus, $\omega_{3}=e^{\sqrt{-1}\frac{2\pi}{3}}$ and the clock operation is $\mathsf{Z}=\ket{0}\bra{0}+\omega\ket{0}\bra{0}+\omega^2\ket{0}\bra{0}$. Then, each $\ket{\phi_m}$, $m \in [3]$, can be written as
\begin{align}
    \ket{\phi_0}&=\frac{1}{\sqrt{3}}\left(\ket{000}+\ket{111}+\ket{222}\right),\\
    \ket{\phi_1}&=\frac{1}{\sqrt{3}}\left(\ket{000}+\omega\ket{111}+\omega^2\ket{222}\right),\\
    \ket{\phi_2}&=\frac{1}{\sqrt{3}}\left(\ket{000}+\omega^2\ket{111}+\omega^4\ket{222}\right).
\end{align}

The shared quantum state of the parties is then given by $\ket{\phi_0}\otimes\ket{\phi_0}\otimes\ket{\phi_0}\otimes\ket{\phi_0}$, i.e., a $\ket{\phi_0}$ state for each element of $\mathcal{K}$. As stated, any arbitrary party can be chosen as the leader, thus without loss of generality, let $L=1$. For security against the eavesdropper, parties $0$ and $2$ will each share a random vector with the leader party. Thus, party $0$ will share $U_0=(U_{00},U_{01},U_{02},U_{03}) \in \mathbb{F}_3^4$ and party $2$ will share $U_2=(U_{20},U_{21},U_{22},U_{23}) \in \mathbb{F}_3^4$ with the leader, prior to the initialization of encoding and after the selection of the leader.

Then, encoding will start. Per Remark \ref{leaderno}, leader does not have to do any encoding operation as it already knows its own incidence vector, but for the sake of the argument, let us consider that it does and as mentioned in Remark~\ref{leaderno}, will not use shared randomness in the encoding stage.

For $a \in \mathcal{K}$, the encoding is given by the operation $\mathsf{Z}^{U_{00}+1}\otimes\mathsf{Z}\otimes \mathsf{Z}^{U_{20}}$. Applying this operation to $\ket{\phi_0}$ gives the result
\begin{align}
    &\!\!\!\!\left(\mathsf{Z}^{U_{00}+1}\otimes\mathsf{Z}\otimes \mathsf{Z}^{U_{20}}\right)\ket{\phi_0}\notag\\
    &=\frac{1}{\sqrt{3}}(\ket{000}+\omega^{2+U_{00}+U_{20}}\ket{111}
    +\omega^{4+2U_{00}+2U_{20}}\ket{222}), \nonumber\\
    &=\ket{\phi_{2+U_{00}+U_{20}}}.
\end{align}
Similar analysis can be carried out for other elements in $\mathcal{K}$. After the encoding stage is done by each party, the final encoded shared quantum state is given by,
\begin{align}
    \ket{\psi_e}=&\ket{\phi_{2+U_{00}+U_{20}}}\otimes\ket{\phi_{1+U_{01}+U_{21}}}\otimes \notag\\&\ket{\phi_{3+U_{02}+U_{22}}} \otimes\ket{\phi_{U_{03}+U_{23}}}.
\end{align}
Afterwards, parties $0$ and $2$ send their shares of qudits to party $1$ for decoding. Using the shared randomness, the leader applies \eqref{decode_step} and obtains,
\begin{align}
\label{finalstate}
    \ket{\psi_f}=\ket{\phi_{2}}\otimes\ket{\phi_{1}}\otimes \ket{\phi_{0}}\otimes\ket{\phi_{0}},
\end{align}
since $3=0$ in $\mathbb{F}_3$. Then, the leader party applies a partial PVM which includes $\ket{\phi_0},\ket{\phi_1},\ket{\phi_2}$ states to each $3$ qudit parts of the tensor product in \eqref{finalstate}. The measurement results will give $2$ for $a$, $1$ for $b$, $0$ for $c$, and $0$ for $d$. However, since the leader already knows it has $c \in E_1$, it can then infer that $c$ is in fact present in every party's subset.

\section{Proofs}\label{proofs}
\begin{lemma}
    The proposed scheme is secure against an eavesdropper.
\end{lemma}

\begin{Proof}
    Note that the $k$th party does the $\mathsf Z$ operation on one of its shared $l$th qudit $(U_k+E_k)_l$ times. Using one-time pad theorem \cite{shannon}, we see that $(E_k)_l$ is secure against a party that does not know $(U_k)_l$. Since this is correct $\forall l \in [K],\ \forall k\in [N]$, the scheme is secure against an eavesdropper.
\end{Proof}

\begin{lemma}
\label{pvm}
    It is possible to construct a partial PVM using $\ket{\phi_m}=\frac{1}{\sqrt{P}}\sum_{k=0}^{P-1}\omega_P^{mk}\underbrace{\ket{k\ldots k}}_{\text{$N$ times}}$ states.
\end{lemma}

\begin{Proof}
    First, note that,
    \begin{align}
        \bra{\phi_m}\ket{\phi_n}&=\frac{1}{N}\sum_{k=0}^{P-1}\sum_{l=0}^{P-1}\omega_P^{-mk}\omega_P^{nl}\bra{k\ldots k}\ket{l\ldots l} \\
        &=\frac{1}{N}\sum_{k=0}^{P-1}\sum_{l=0}^{P-1}\omega_P^{nl-mk}\delta_{k,l} \\
        &=\frac{1}{N}\sum_{k=0}^{P-1}\omega_P^{(n-m)k} \\
        &=\frac{1}{N}\frac{1-\omega_P^{(n-m)P}}{1-\omega_P^{(n-m)}}\\
       \label{fouriereq} &=\delta_{n,m}
    \end{align}
    Notice that $\underbrace{\ket{k\ldots k}}_{\text{$N$ times}}$ is $P^N$ dimensional vector. Hence, it commits a basis of $P^N$ vectors. Since $N \geq 1,\ P \geq N$, we have $P^N \geq N-1$. Thus, using $\ket{\phi_m}$ with $m \in [N]$, the leader party can always construct an orthonormal basis including these states, possibly using a procedure such as Gram-Schmidt.

    Then, using that orthonormal basis and invoking Lemma~\ref{meas}, the leader can construct a PVM $\{P_i\}_i$ for a $P^N$ dimensional Hilbert space. Then, this PVM can be used to construct the partial PVM $\{P_i\otimes\underbrace{I_{P}\otimes\ldots\otimes I_{P}}_{\text{$N-1$ times}}\}_i$.

    This process can then be used similarly to construct a partial PVM for the other $N$-qudit parts of the $K$-separation mentioned in Section V.
\end{Proof}

\begin{remark}
    From \eqref{fouriereq} and the superposition coefficients of $\ket{\phi_m}$ states in computational basis, it can be observed that $\ket{\phi_m}$ states are $N$-qudit Fourier states for the $P$-dimensional subspace of $\mathcal{H}^{
    \otimes N}$.
\end{remark}

\begin{lemma} \label{priv}
    The proposed scheme satisfies the privacy criterion.
\end{lemma}

\begin{Proof}
    Note that,
    \begin{align}
        \omega_P^k\underbrace{\ket{k\ldots k}}_{\text{$N$ times}}&=(\mathsf{Z}\otimes \underbrace{I_{P}\otimes\ldots I_P}_{\text{$N-1$ times}})\underbrace{\ket{k\ldots k}}_{\text{$N$ times}}\\
        &=(I_P\otimes \mathsf{Z}\otimes \underbrace{I_{P}\otimes\ldots I_P}_{\text{$N-2$ times}})\underbrace{\ket{k\ldots k}}_{\text{$N$ times}}\\
        \label{blind}&=(\underbrace{I_P\otimes\ldots\otimes I_P}_{\text{$l$ times}}\otimes \mathsf{Z}\otimes \underbrace{I_{P}\otimes\ldots I_P}_{\text{$N-l-1$ times}})\underbrace{\ket{k\ldots k}}_{\text{$N$ times}}
    \end{align}
    Thus, any linear combination of $\underbrace{\ket{k\ldots k}}_{\text{$N$ times}}$ states will be blind to which party applied the $\mathsf{Z}$ operation.

    Then, using the fact that $(A\otimes B)=(A\otimes I)(I\otimes B)$, we see that privacy is guaranteed no matter how many parties apply the $\mathsf{Z}$ operation.

    From \eqref{blind}, we note that the posterior estimation of the leader party after the scheme about $(E_k)_l$ is,
    \begin{align}
        P((E_k)_l=1&|A_{[N]}= \ket{\phi_m}) \nonumber\\
        &=\frac{P((E_k)_l=1,A_{[N]}=\ket{\phi_m})}{P(A_{[N]}=\ket{\phi_m})}\\
        &=\frac{P((E_k)_l=1,\sum_{i=0}^{N-1}(E_i)_l=m)}{P(\sum_{i=0}^{N-1}(E_i)_l=m)}\\
        &=P((E_k)_l=1|\sum_{i=0}^{N-1}(E_i)_l=m)),
    \end{align}
    as $\ket{\phi_m}$ can happen with any $m$ parties having $1$ at the $l$th index of their incidence vector. Thus, from the answers, the leader cannot learn anything more than what could be learnt from $\sum_{i=0}^{N-1}E_i$. Hence, the privacy constraint is satisfied.
\end{Proof}

\begin{lemma}\label{correct}
    The proposed scheme satisfies the correctness criterion.
\end{lemma}

\begin{Proof}
    From \eqref{blind} and the fact that $(A\otimes B)=(A\otimes I)(I\otimes B)$, if $m$ parties apply the $\mathsf Z$ operation to their qudits, 
    \begin{align}
        (I_P\otimes&\ldots \otimes \prescript{i(0)}{}{\mathsf{Z}}\otimes\ldots\otimes \prescript{i(m-1)}{}{\mathsf{Z}}\otimes I_P)\underbrace{\ket{k\ldots k}}_{\text{$N$ times}} \notag\\ \label{sumpriv}
        &=(\underbrace{I_p\ldots \otimes I_P}_{\text{$k$ times}}  \otimes \mathsf{Z}^m\otimes\underbrace{I_p\ldots \otimes I_P}_{\text{$N-k-1$ times}})\underbrace{\ket{k\ldots k}}_{\text{$N$ times}},
    \end{align}
    where the left superscript indicates which party does the $\mathsf Z$ operation and $i:[m]\to [N]$ is an injection; see Remark~\ref{distribution}. Then, we have,
    \begin{align}
        (\mathsf{Z}^m\otimes \underbrace{I_P\otimes\ldots\otimes I_P}_{\text{$N-1$ times}})\ket{\psi}&=\frac{1}{\sqrt{P}} (\mathsf{Z}^m\otimes\underbrace{I_P\otimes\ldots\otimes I_P}_{\text{$N-1$ times}})\notag\\ &\quad\quad\quad \sum_{n=0}^{P-1}\underbrace{\ket{n\ldots n}}_{\text{$N$ times}} \\
        &=\frac{1}{\sqrt{P}}\sum_{n=0}^{P-1}\omega_P^{mn}\underbrace{\ket{n\ldots n}}_{\text{$N$ times}} \\
        &=\ket{\phi_m}.
    \end{align}

    Thus, using the partial PVMs constructed in Lemma~\ref{pvm}, the leader party can carry out partial measurements. If the result is one of the $\ket{\phi_m}$ states for $m\in [N]$, it means that $m$ parties applied the $\mathsf Z$ operation, thus the membership aggregation result is $m$. However, as mentioned in Lemma~\ref{pvm}, since there are at least $N$ elements in that PVM, if anything other than a $\ket{\phi_m}$ has been measured, it indicates the existence of a Byzantine element in the system.
\end{Proof}

\begin{remark}
    We note here that although the scheme sometimes detects the Byzantine elements, it may fail to detect them as well. Moreover, even in the cases it detects them, it cannot correct them. Thus, the scheme is not Byzantine-proof.
\end{remark}

\begin{remark}
    If $N$ is a prime number itself, then $P=N$ is a valid choice for the underlying field. However, note that, in that case, $N$ parties applying the $\mathsf Z$ operation is equivalent to no parties applying it. Yet, the leader is able to differentiate between those cases based on its set membership result.
\end{remark}

\begin{remark}
\label{distribution}
     Equation \eqref{sumpriv} can actually be generalized. Let $f:[N] \to \mathbb{F}_p$ such that $\sum_{k\in [N]}f(k)=m$. Then,
    \begin{align}
        (\mathsf{Z}^{f(0)}&\otimes\ldots \otimes \mathsf{Z}^{f(N-1)})\underbrace{\ket{k\ldots k}}_{\text{$N$ times}} \notag\\
        &=(\underbrace{I_p\ldots \otimes I_P}_{\text{$k$ times}}  \otimes \mathsf{Z}^m\otimes\underbrace{I_p\ldots \otimes I_P}_{\text{$N-k-1$ times}})\underbrace{\ket{k\ldots k}}_{\text{$N$ times}},
    \end{align}
\end{remark}

\begin{corollary}
   Looking at Lemma~\ref{pvm} and Remark~\ref{distribution}, if the partial PVMs include $\ket{\phi_m}$ for $m \in [P]$, the proposed scheme carries out quantum private summation modulo $P$.
\end{corollary}

\section{Conclusions}\label{conclusions} 
We developed a private membership aggregation scheme using quantum states. The scheme provides privacy to the legitimate parties that participate in the scheme, and security against external eavesdroppers. Essentially, the proposed scheme is a quantum private summation modulo $P$ scheme. The proposed scheme makes use of the phase operation to encode the information into specific $N$-qudit Fourier transform states, which then makes use of the fact that the Fourier transform is a unitary operation to construct an orthonormal measurement basis. However, one caveat of using such bases is the fact that the Fourier transform is only a unitary operation on a field, as it involves addition and multiplication. Thus far, the results in the literature lack other parallel-operating private quantum summation algorithms that carry out the private summation on ring or even group structures rather than relying on fields, which can be the point of a future study. 

\newpage
\IEEEtriggeratref{15}
\bibliographystyle{unsrt}
\bibliography{references}
\end{document}